\documentclass[aps,prd,twocolumn,showpacs,superscriptaddress,nofootinbib,preprintnumbers]{revtex4-1}

\usepackage{style}

\begin{document}

\preprint{FERMILAB-PUB-24-0744-T}

\title{The Affleck-Dine Curvaton}

\author{Aurora Ireland\,\orcidlink{0000-0001-5393-0971}} 
\email{anireland@stanford.edu}
\affiliation{Stanford Institute for Theoretical Physics, Department of Physics, Stanford University, Stanford, CA 94305}

\author{Gordan Krnjaic\,\orcidlink{0000-0001-7420-9577}} 
\email{krnjaicg@uchicago.edu}
\affiliation{Theoretical Physics Division, Fermi National Accelerator Laboratory, Batavia, Illinois 60510}
\affiliation{Kavli Institute for Cosmological Physics, University of Chicago, Chicago, IL 60637}
\affiliation{Department of Astronomy and Astrophysics, University of Chicago, Chicago, IL 60637}

\author{Takuya Okawa\,\orcidlink{0009-0003-9478-380X}}
\email{o.takuya@wustl.edu}
\affiliation{Theoretical Physics Division, Fermi National Accelerator Laboratory, Batavia, Illinois 60510}

\affiliation{Physics Department and McDonnell Center for the Space Sciences, Washington University in St. Louis, St. Louis, MO 63130}

\date{\today}

\begin{abstract}




The Standard Model of particle physics does not explain the origin of the universe's baryon asymmetry or its primordial fluctuations. 
The Affleck-Dine mechanism is a well motivated scenario for generating the baryon asymmetry through the post-inflationary dynamics of a complex scalar field with baryon number.
The curvaton mechanism is a popular approach for producing curvature perturbations through the dynamics of a light spectator field which decays after inflation.
We demonstrate that the same complex field can viably perform both roles without any modifications to the minimal realization of Affleck-Dine baryogenesis. This scenario can also accommodate appreciable levels of primordial non-Gaussianity, beyond those achievable with only a real-valued curvaton field, and may be observable with future CMB experiments.

\end{abstract}
\bigskip
\maketitle

\section{Introduction}\label{sec:intro}

Our universe exhibits nearly scale invariant perturbations that seed  cosmic structures and imprint temperature anisotropies onto the cosmic microwave background (CMB) \cite{Planck:2018vyg}. Since these fluctuations are correlated on scales outside the cosmological horizon during recombination, 
they can be dynamically generated 
during a period of inflation  when the comoving Hubble radius decreases with time and modes become frozen on superhorizon scales. 
When inflation ends and the horizon begins to grow, they eventually re-establish causal contact, but retain large scale correlations --- see Ref. \cite{Ellis:2023wic} for a review. Since the Standard Model (SM) does not explain the origin of these fluctuations, their existence is technically evidence of new physics.\footnote{In principle, the SM Higgs field could drive inflation and generate the curvature perturbation, but viable realizations require e.g. non-minimal couplings to gravity \cite{Rubio:2018ogq}.}

Although it is often assumed that these fluctuations arise from the field driving inflation, this need not be the case. Curvaton models \cite{Lyth:2001nq,Enqvist:2001zp,Bartolo:2002vf} posit an additional light scalar field which is a spectator during inflation, during which it acquires isocurvature perturbations. After reheating, the curvaton decays to radiation, and its isocurvature perturbations are converted to adiabatic. The result is generically larger primordial non-Gaussianities, as compared with models where a single inflaton generates perturbations during slow-roll evolution.

Generating the baryon asymmetry of the universe also necessitates new physics \cite{Sakharov:1967dj}. Although the SM does include baryon number violating and both $C$- and $CP$-violating interactions, the amounts of violation are insufficient to explain the observed asymmetry. Further, these processes must occur out-of-equilibrium. While the electroweak phase transition could in principle provide such conditions, extensions are necessary to make it strongly first order \cite{Morrissey:2012db}. As it stands, the SM does not provide a complete framework to account for the observed baryon-to-photon ratio $\eta_B = n_B/n_\gamma  = 6 \times 10^{-10}$ \cite{Planck:2018vyg} --- see Ref. \cite{Cline:2006ts} for a review. 

A popular scenario for generating the baryon asymmetry is the Affleck-Dine mechanism, in which a complex scalar field with non-zero baryon number is initially displaced from the minimum of its potential in the early universe \cite{Affleck:1984fy}. If the potential contains feeble baryon violating self-interactions, as the scalar evolves the comoving baryon number begins to accumulate until the scalar decays to SM particles through baryon preserving operators to transfer this asymmetry to the SM plasma. Unlike most other mechanisms, Affleck-Dine stands out for generically predicting too large of an asymmetry unless the parameters that govern the baryon density have extremely small values.

In this {\it Letter}, we demonstrate that the same complex scalar field realizing Affleck-Dine baryogenesis can also serve as the curvaton without sacrificing any of the appealing features of either mechanism.\footnote{It has previously been shown that the inflaton can double as the Affleck-Dine field in certain inflationary contexts --- see e.g. Refs.~\cite{Charng:2008ke,Hertzberg:2013mba,Cline:2019fxx,Lin:2020lmr}. The advantage of our Affleck-Dine curvaton scenario is its minimality and negligible fine tuning.} Specifically, we find that within this combined framework, the observed baryon asymmetry can be generated with the same model parameters that also yield the observed curvature perturbation. A distinctive feature of this scenario is larger levels of primordial non-Gaussianity, which can in principle be observed in future CMB experiments.


\section{Affleck-Dine Review}\label{sec:Affleck-Dine}

We begin by briefly reviewing the Affleck-Dine mechanism for generating the baryon asymmetry \cite{Affleck:1984fy}. Consider extending the SM to include a complex scalar field $\chi$ with non-zero baryon number $B_\chi$.\footnote{This is a well-motivated scenario which arises in many supersymmetric extensions to the SM \cite{Enqvist:2003}.} It is useful to decompose $\chi$ into a modulus $r$ and complex phase $\theta$,
\begin{align}
    \chi = \frac{r}{\sqrt{2}} e^{i \theta} \,.
\end{align}
Under global phase rotation, $\chi$ transforms as $\chi \rightarrow e^{i\alpha B_\chi} \chi$, with $\alpha$ an arbitrary constant. The baryon number density stored in the $\chi$ field is identified with the time-like component of the baryon Noether current 
\begin{align}
\label{eq:nB}
    n_B= i B_\chi (\dot{\chi}^*\chi - \chi^* \dot{\chi}) = B_\chi r^2 \dot{\theta} \,,
\end{align}
which is only non-zero when the complex phase is dynamical in the early universe. 

The potential for $\chi$ has both baryon preserving and (small) baryon violating self interactions 
\be
V(\chi) =  m_\chi^2 |\chi|^2 + \frac{\lambda}{2}  |\chi|^4 + \frac{ \lambda^\prime }{4} (\chi^4 + \chi^{*4} ) \,,
\ee
where the last term explicitly breaks baryon number, so we demand $\lambda^\prime \ll \lambda$. In terms of $r$ and $\theta$, the action is
\begin{align}
    S = \int d^4x   \, a^3 \!\left( \frac{1}{2} \dot{r}^2+\frac{1}{2} r^2 \dot{\theta}^2-V(r, \theta) \right),
\end{align}
where $a$ is the scale factor. The radial equation of motion can be written
\be
\label{eq:eom-r}
   \ddot{r} + 3H \dot{r} - \dot{\theta}^2 r  = - \frac{\partial V}{\partial r} \approx - m_\chi^2 r  \,,
\ee
where $H = \dot a/a$ is the Hubble rate and for simplicity we have assumed $\chi \ll m_\chi/\sqrt{\lambda}$ such that the potential is always dominated by the quadratic interaction.\footnote{See e.g. Refs.~\cite{Enqvist:2009ww,Enqvist:2009zf,Enqvist:2010ky,Hooper:2023nnl} for model-building efforts with a self-interacting curvaton.}
Using \Eq{eq:nB}, the equation of motion for $\theta$ can be written in terms of the  baryon number density
\begin{align}
\label{eq:nB-EOM}
\frac{1}{a^3}   \frac{\partial }{\partial t} \! \left(a^3 n_B \right) = - \frac{ \lambda^\prime B_\chi}{2} \, r^4 \sin(4 \theta)~.
\end{align}
If $\lambda^\prime = 0$, co-moving baryon number is conserved, in analogy with angular momentum conservation
in classical systems with radially symmetric potentials.
Integrating \Eq{eq:nB-EOM} with the symmetric initial condition $n_B(t_i) = 0$, we obtain 
\begin{align}
    \label{eq:baryon number time evolution}
     n_B(t) = \frac{ \lambda^{\prime} B_\chi }{2 a^3(t)} \int_{t_i}^t  dt^{\prime} a^3\! \left(t^{\prime}\right) r^4 \!\left(t^{\prime}\right)  \sin [4 \theta\!\left(t^{\prime}\right) ] \,,
\end{align}
which is valid for $\tau_\chi > t > t_i$. Here, $\tau_\chi$ is the $\chi$ lifetime and $t_i$ is the initial time, which we identify with reheating.

In order to transfer the asymmetry, $\chi$ must also couple to SM fields through baryon number {\it preserving } interactions of the form 
 \be 
 \label{eq:lint}
 {\cal L}_{\rm int} =  c {\cal \hat O}_{\rm SM} \chi + h.c.~,
 \ee
where ${\cal \hat O}_{\rm SM}$ is a composite operator of SM fields with baryon number $B_{\cal \hat O} =  - B_\chi$ and $c$ is a constant coefficient.\footnote{For additional flexibility, it is often advantageous to also introduce a new fermion $\psi$ which does not carry baryon number, such that one can write a Yukawa term of the form $\mathcal{L}_{\rm int} = y \bar{q} \psi \chi$, where $q$ may be a composite operator of SM fields carrying baryon number $B_q = - B_\chi$.} Thus, viable baryogenesis can achieved if $\chi$ decays to SM fields through the interaction in \Eq{eq:lint} before the asymmetry is depleted in the early universe. If there are no other decay channels for $\chi$, the baryon-to-photon ratio is 
\be
\eta_B = 
\frac{\pi^2 }{4 \zeta(3) }
\frac{\lambda^{\prime} B_\chi }{T^3_{ i}   } \int_{t_i}^{\tau_\chi}  dt^{\prime} a^3\! \left(t^{\prime}\right) r^4 \!\left(t^{\prime}\right)  \sin [4 \theta\!\left(t^{\prime}\right) ] \,,
\ee
where  $T_i \equiv T_\gamma(t_i)$ is the  initial photon temperature. In our later analysis, we take $B_\chi =1$ without loss of generality.

\section{Curvaton Review}\label{sec:curv}

In the original curvaton scenario \cite{Lyth:2001nq,Enqvist:2001zp,Bartolo:2002vf}, a real scalar field $\sigma$ is present as a light spectator during inflation. The energy density of the curvaton is always subdominant during inflation, such that it does not drive Hubble expansion; it is, however, responsible for eventually generating the dominant contribution to the curvature perturbation. The minimal realization posits a quadratic background potential $V(\sigma) = \frac{1}{2} m_\sigma^2 \sigma^2$ with $m_\sigma \ll H_i$, such that the curvaton acquires Gaussian isocurvature fluctuations of amplitude $\delta \sigma_i \sim H_i/(2\pi)$, which are the dominant influence on its evolution. Here $H = \dot a /a$ is the Hubble rate and the subscript $i$ denotes a value during inflation. 

When inflation ends, the inflaton decays to reheat the universe. The curvaton is initially Hubble damped, but begins to oscillate about its quadratic potential when $H \sim m_\sigma$. During this oscillatory phase, the curvaton redshifts like nonrelativistic matter $\rho_\sigma \propto a^{-3}$ and the energy density in $\sigma$ grows relative to that in radiation. Finally, when  $H \sim \tau_\sigma^{-1}$, where $\tau_\sigma$ is the curvaton lifetime, $\sigma$ decays to SM radiation and its isocurvature perturbations are converted to adiabatic perturbations, presuming the decay products thermalize with the background plasma.\footnote{If dark matter has been produced by this point and is decoupled from the radiation, then curvaton decays induce a mismatch between photon and dark matter perturbations, which can conflict with CMB measurements. It is straightforward to avoid this fate (e.g. by producing dark matter thermally after the curvaton decays), so we ignore this possibility.}

\begin{figure}[t!]
\hspace{-1cm}
    \includegraphics[width=\linewidth]{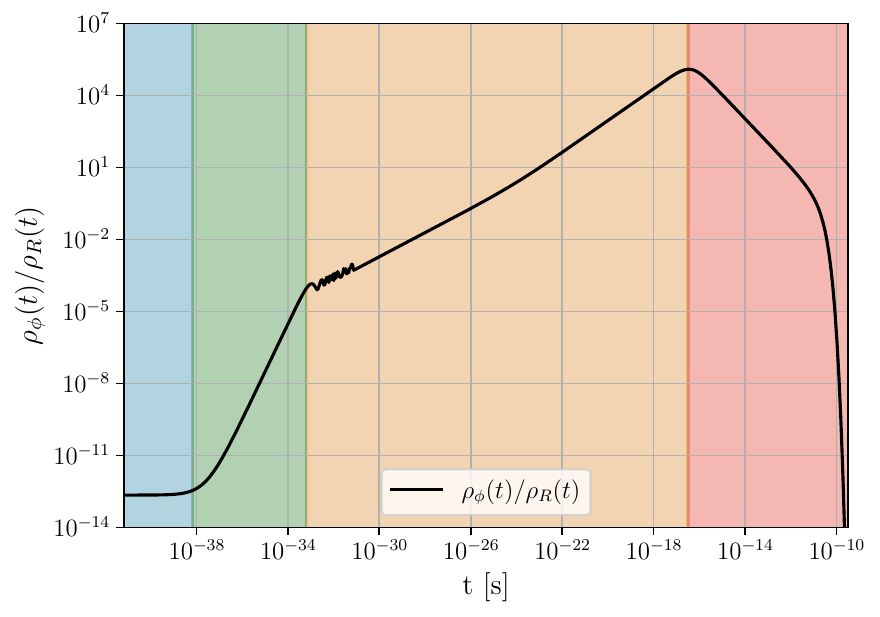}
    \caption{The curvaton-radiation energy density ratio in various cosmic phases: the Hubble-damped regime  (green), the quadratic oscillation regime (orange), and the $\chi$ decay regime (red). Here we assume the benchmark parameters from Model 1 in Table~\ref{tab:successful model}.
     }
    \label{fig:energy density ratio}
\end{figure}

To calculate $\zeta$ in this scenario, it is useful to employ the $\delta N$ formalism \cite{Salopek:1990,Starobinsky:1982,Sasaki:1995,Sasaki:1998,Lyth:2004gb}, in which the curvaton curvature perturbation $\zeta_\sigma$ on a general hypersurface of uniform curvaton density is \cite{Sasaki:2006kq}
\be
\label{eq:zeta_sigma_deltaN}
\zeta_\sigma(t, \vec x) = \delta N(t, \vec{x}) +  \frac{1}{3} \log\brac{\rho_{\sigma}(t, \vec{x})}{\bar \rho_\sigma(t)} \,,
\ee
where $\delta N$ is the perturbed number of $e$-folds, $\rho_\sigma$ is the curvaton energy density,
and bars denote background quantities. On a uniform \textit{total} density slice, we can equate $\delta N$ with the total curvaton perturbation $\zeta = \delta N$ and write
\be
\rho_\sigma(t, \vec{x}) = e^{3(\zeta_\sigma - \zeta)} \bar \rho_\sigma(t)~~,~~ 
\rho_R(t, \vec{x}) \approx e^{-4 \zeta} \bar \rho_R(t),~ 
\ee
where $\rho_R$ is the energy density of SM radiation and we have assumed $\zeta_\sigma \gg \zeta_R$.
At the time of curvaton decay, energy conservation yields an implicit relation for the curvature perturbation 
\be
\label{eq:implicit}
e^{4\zeta} - \left( e^{3\zeta_\sigma} \Omega_{\sigma} \right)e^\zeta + (\Omega_{\sigma} - 1) = 0,
\ee 
where $\Omega_i \equiv \bar \rho_i(\tau_\sigma)/\bar \rho_{\rm tot}(\tau_\sigma)$, $\bar \rho_{\rm tot} = \bar{\rho}_\sigma + \bar{\rho}_R$ is the total background energy density, and we have assumed that total density is uniform on the decay surface. In terms of the curvaton energy fraction at decay
$\kappa  = 3 \Omega_{\sigma,{\tau_\sigma}}/(4 - \Omega_{\sigma,{\tau_\sigma}})$,
the general solution is 
\begin{equation}
\label{eq:lnX}
    \zeta = \log \left( \frac{  \beta^{1/2} + \sqrt{  \alpha \kappa \beta^{-1/2} - \beta}}{(3+\kappa)^{1/3}}  \right) \,,
\end{equation}
where we have defined parameters
\be
    \beta &=& \frac{1}{2} \left[ \gamma^{1/3} + (\kappa-1)(\kappa+3)^{1/3} \gamma^{-1/3} \right]  \\[6pt] 
    \gamma &=& (\alpha \kappa)^2 + \sqrt{(\alpha \kappa)^4 + (3+\kappa)(1-\kappa)^3} \, ,
\ee
and for a strictly quadratic potential $\zeta_\sigma$ can be related to the initial field perturbation as \cite{Sasaki:2006kq}
\be
\alpha \equiv e^{3\zeta_\sigma} = \left( 1 + \frac{\delta \sigma_i}{\bar \sigma_i} \right)^2,
\ee
where $\bar\sigma_i$ and $\delta \sigma_i$ are the  curvaton background value and perturbation during inflation, respectively. Thus, $\zeta$ depends only on the initial curvaton perturbation and the curvaton energy fraction at the time of decay. 

The expression in \Eq{eq:lnX} is a non-linear mapping between the Gaussian perturbation $\zeta_\sigma$ and the total curvature perturbation $\zeta$, so this scenario generically yields larger non-Gaussianities than models where a minimal inflaton generates all perturbations. Taylor expanding in the Gaussian variable $\zeta_g$ to third order yields 
\be
\label{eq:ng-expansion}
    \zeta \simeq \zeta_\sigma + \frac{3}{5} f_{\rm NL} \zeta_\sigma^2 + \frac{9}{25} g_{\rm NL} \zeta_\sigma^3 \,,
\ee
where we have defined the non-linearity paramters $f_{\rm NL}$ and $g_{\rm NL}$, constrained by the Planck collaboration as \cite{Planck:2019kim}
\be
    f_\mathrm{NL} = 5.0 \pm 8.4 ~~,~~
    g_\mathrm{NL} = (-5.8 \pm 6.5) \times 10^4~,
\ee
which are both consistent with zero. For a discussion of primordial non-Gaussianity in curvaton models, see Ref. \cite{Bartolo:2002vf}. We now turn to generalizing this argument in the case of a complex curvaton identified with the Affleck-Dine scalar.





\section{Combined Scenario}\label{sec:combined}

We now promote the original Affleck-Dine field $\chi$ from Sec \ref{sec:Affleck-Dine} to also play the role of the curvaton. The key conceptual difference in this scenario 
is that $\chi = \chi_R + i \chi_I$ is a complex field with two degrees of freedom
\be
\label{eq:re-im}
r = \sqrt{2( \chi_R^2 + \chi_I^2 ) }~~,~~ \theta \equiv \tan^{-1} \brac{\chi_I}{\chi_R},
\ee
which are related through the equations of motion in \Eq{eq:eom-r} and \Eq{eq:nB-EOM}. In our numerical analysis, these equations are supplemented with decay terms and in
Fig. \ref{fig:energy density ratio} we show the $\chi$ energy density as a function of time for a benchmark model.

\subsection{Curvature Perturbation}
Since $m_\chi \ll H_i$, the field components $\chi_{R,I}$ acquire uncorrelated quantum fluctuations $\delta \chi_R$ and $\delta {\chi_I}$ during inflation, which evolve after inflation to set the values of the corresponding curvature perturbations $\zeta_{\chi_R}$ and $\zeta_{\chi_I}$. The curvature perturbation from two independent curvaton fields has been studied in Ref.~\cite{Assadullahi:2007uw}, and we use this formalism to describe our scenario in which the dynamics of $\chi_R$ and $\chi_I$ are related through the role they play in setting the complex phase in \Eq{eq:re-im}.

\begin{table}[t]
    \centering
    \begin{tabular}{|c|c|c|c|}
        \hline
        & Model 1 & Model 2 & Model 3 \\ \hline 
        \\[-1em]
        $m_\chi/{\rm GeV}$ & $10^{9}$ & $10^{10}$ & $10^{11}$ \\
        $\Gamma_\chi/\mathrm{GeV}$ & $10^{-13}$ & $10^{-13}$ & $10^{-13}$ \\
        $H_i/M_\mathrm{pl}$ & $1 \times 10^{-6} $ & $1 \times 10^{-7} $ & $2 \times 10^{-9} $ \\
        $r_i/H_i$ & $1 \times 10^4$ & $5 \times 10^3$ & $10^4$ \\
        $\theta_i/(\pi/8)$ & $1.1$ & $1.1$ & $1.1$ \\
        $\lambda$ & $10^{-20}$ & $10^{-20}$ & $10^{-10}$ \\
        $\lambda^\prime$ & $10^{-24}$ & $10^{-17}$ & $10^{-9}$ \\
        \hline 
        \\[-1em]
        $\zeta~[10^{-5}]$ & $-2.3$ & $4.6$ & $2.2$ \\
        $f_\mathrm{NL}^\mathrm{local}$ & $-0.88$ & $-0.88$ & $-0.87$ \\
        $\eta_B~[10^{-10}]$ & $4.9$ & $2.6$ & $3.1$ \\
        $r_T$ & $1.1 \times 10^{-3}$ & $3.0 \times 10^{-6}$ & $5.1 \times 10^{-9}$ \\
        $\alpha_{II}$ & $1.1 \times 10^{-3}$  & $7.4 \times 10^{-4}$ & $7.8 \times 10^{-4}$  \\
    \hline
    \end{tabular}
    \caption{Three benchmark models which predict acceptable values for $\zeta$, the local non-linearity parameter $f_{\rm NL}^{\rm local}$, $\eta_B$, the tensor-to-scalar ratio $r_T$, and $\alpha_{II}$.}
    \label{tab:successful model}
\end{table}

Repeating the same sequence of steps that leads to \Eq{eq:implicit},  for a complex curvaton we find
\begin{align}
    \Omega_\gamma e^{4\left(\zeta_{\gamma}-\zeta\right)} + \Omega_{\chi_R} e^{3\left(\zeta_{\chi_R}-\zeta\right)} + \Omega_{\chi_I} e^{3\left(\zeta_{\chi_I}-\zeta\right)} &= 1 \,,
    \label{eq:density contrast relations 1}
\end{align}
where $\Omega_{i}=\bar{\rho}_{i}(\tau_\chi) / \bar{\rho}_{\rm tot}(\tau_\chi)$  are the density parameters of each field evaluated at the time of $\chi$ decay. We numerically evolve this system to obtain the parameters $\Omega_i$ and then solve for the perturbation of each field to determine $\zeta$ using \Eq{eq:density contrast relations 1}.
Taylor expanding Eq.~(\ref{eq:density contrast relations 1}) to first order in all perturbations, we find 
\be
    4 \Omega_\gamma (\zeta_{\gamma} - \zeta) + 3 \Omega_{\chi_R}(\zeta_{\chi_R} - \zeta)
    + 3 \Omega_{\chi_I} (\zeta_{\chi_I} - \zeta) = 0 ,~
\ee
and the linear-order solution for the total curvature perturbation is then
\be
\label{eq:zeta-final}
    \zeta 
    = f_\gamma \zeta_{\gamma} + f_{\chi_R}\zeta_{\chi_R} + f_{\chi_I} \zeta_{\chi_I} \,, 
\ee
where we have defined the parameters 
\begin{align}\label{eq:fs}
    f_\gamma &= \frac{4 \Omega_\gamma}{4 \Omega_\gamma + 3 \Omega_{\chi_R} + 3 \Omega_{\chi_I}} \,, \\
    f_{\chi_R} &= \frac{3 \Omega_{\chi_R}}{4 \Omega_\gamma + 3 \Omega_{\chi_R} + 3 \Omega_{\chi_I}} \,, \\
    f_{\chi_I} &= \frac{3 \Omega_{\chi_I}}{4 \Omega_\gamma + 3 \Omega_{\chi_R} + 3 \Omega_{\chi_I}} \,.
\end{align}
Using \Eq{eq:zeta-final}, we can obtain the final curvature perturbation at the time of $\chi$ decay. Note, however, that these relations only hold at linear order, and calculating the primordial non-Gaussianity parameters --- defined in \Eq{eq:ng-expansion} --- requires a higher order expansion. Thus in Appendix \ref{sec:second}, we present the second order relations we use to compute $f_{\rm NL}$ in our analysis.

\subsection{Non-Gaussianity}

In this subsection, we follow the treatment in Ref.~\cite{Assadullahi:2007uw} to extract $f_{\rm NL}$ in our scenario. To leading order, the primordial curvature perturbation power spectrum is 
\begin{align}
    \label{eq: power spectrum 1}
    P_\zeta = f_{\chi_R}^2 P_{\chi_R} + f_{\chi_I}^2 P_{\chi_I} \,.
\end{align}
The power spectra of the curvaton components at the onset of curvaton oscillations $H \sim m_\chi$ are related by
\be
    \label{eq: power spectrum 2}
    P_{\chi_R} = \beta^2 P_{\chi_I} \,,
\ee
which holds at leading order. In the linear approximation, $\beta$ is reduced to the ratio of the values of the curvaton fields at horizon crossing 
\be
\beta = \frac{  \bar{\chi}_{I}  }{  \bar{\chi}_{R} }= \tan\theta \,,
\ee
where the bar denotes values during inflation. Since the Fourier components of $\chi_R$ and $\chi_I$ are uncorrelated, we have
\be
\left<\zeta_{\chi_R}(\mathbf{k})\zeta_{\chi_I}(\mathbf{k}^\prime)\right> = 0 \,.
\ee
Using also that the correlation function of an odd number of  Gaussian perturbations vanishes, the three-point correlation function of $\zeta$ can be written
\be
\label{eq: three-point correlation function}
    && \hspace{-0.5cm} \left< \zeta(\mathbf{k_1}) \zeta(\mathbf{k_2}) \zeta(\mathbf{k_3}) \right> \nonumber\\
    &=& \frac{1}{2} f_{\chi_R}^2 \left(C - \frac{3}{2}A\right) \left< \zeta_{\chi_R}(\mathbf{k_1}) \zeta_{\chi_R}(\mathbf{k_2}) (\zeta_{\chi_R} * \zeta_{\chi_R})(\mathbf{k_3}) \right> 
    \nonumber\\
    &&  + \frac{1}{2} f_{\chi_I}^2 \left(D - \frac{3}{2}B\right) \left< \zeta_{\chi_I}(\mathbf{k_1}) \zeta_{\chi_I}(\mathbf{k_2}) (\zeta_{\chi_I} * \zeta_{\chi_I})(\mathbf{k_3}) \right> \nonumber \\
    && + \frac{1}{2} f_{\chi_R} f_{\chi_I} E \left< \zeta_{\chi_R}(\mathbf{k_1}) \zeta_{\chi_I}(\mathbf{k_2}) (\zeta_{\chi_R} * \zeta_{\chi_I})(\mathbf{k_3}) \right> \nonumber \\
    &&+ (\mathrm{2~permutation~terms})
\ee
where the parameters $A,B,C,D$ and $E$ are all functions of $f_{\chi_R}, f_{\chi_I}$ and $f_{\gamma}$ defined in Appendix \ref{sec:second}, and an asterisk between functions signifies convolution
\begin{align}
    (\zeta_i * \zeta_j)(\mathbf{k}) =  \int 
    \frac{d^3 q  }{(2\pi)^3}~\zeta_i(\mathbf{q})\zeta_j(\mathbf{k-q}).
\end{align}
Rewriting Eq.~\eqref{eq: three-point correlation function} in terms of the power spectrum of $\zeta$ by using Eqs.~\eqref{eq: power spectrum 1} and~\eqref{eq: power spectrum 2}, one finds
\be
    f_\mathrm{NL} \!= \!\frac{ 5 [\beta^4 f_{\chi_R}^2 \! (2C \!-\! 3A) \!+ \!\beta^2 f_{\chi_R} f_{\chi_I} E + f_{\chi_I}^2 \!(2D \!-\!3B)]}{6(\beta^2 f_{\chi_R}^2 + f_{\chi_I}^2)^2} ,~~
\ee
which we use to obtain our numerical results.

\begin{figure}[t!]
\hspace{-0.5cm}
    \includegraphics[width=0.95\linewidth]{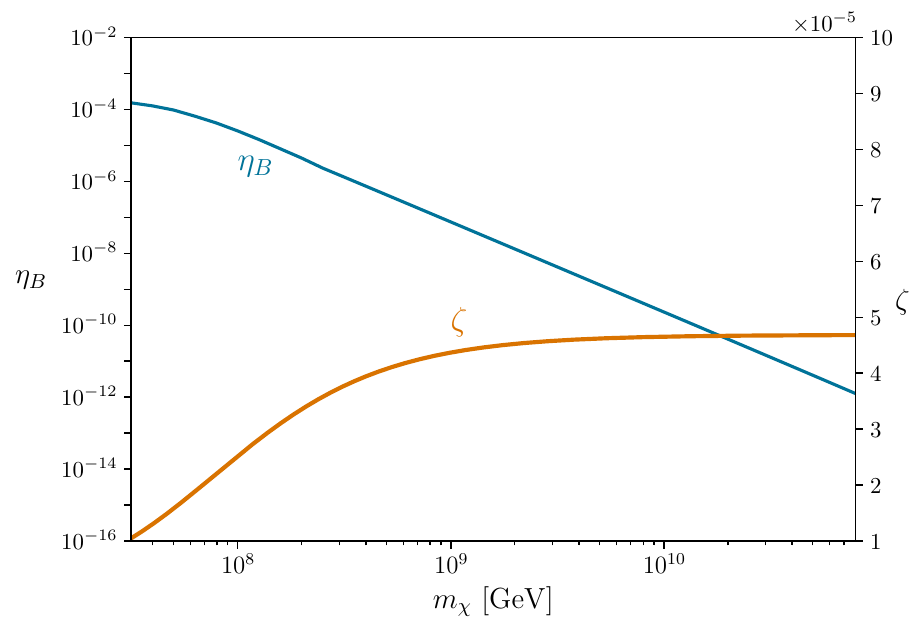}    
    \caption{The baryon-to-photon ratio $\eta_B$ and curvature perturbation $\zeta$ as a function of curvaton mass $m_\chi$. For all points along the $\eta_B$ curve, $\zeta$ is chosen to have the observed value $\zeta \simeq 2 \times 10^{-5}$ and $\lambda^\prime = 10^{-15}$. For all points along the $\zeta$ curve, $\eta_B = 6 \times 10^{-10}$ to match the observed asymmetry, $\lambda^\prime$ is chosen accordingly, and $H_i = 3 \times 10^{-5} M_{\rm pl}$. Unless otherwise specified, all other parameters match the values from Model 2 in Table \ref{tab:successful model}. 
    } 
    \label{fig:results}
\end{figure}

\subsection{Results}
In Table \ref{tab:successful model} we present three benchmark realizations of our scenario with different curvaton masses. In each case, the baryon asymmetry and the primordial curvature perturbation both achieve their observed values. We also show the degree of non-Gaussianity predicted in each scenario and find values of $f_{\rm NL}$ within the forecast sensitivities of CMB-S4 \cite{CMB-S4:2016ple}. However, a complete analysis of the observational consequences is beyond the scope of this work. We have checked also that all benchmark points satisfy the bounds from baryon isocurvature (see supplemental material for detail).

In Fig. \ref{fig:results}, we show how $\eta_B$ and $\zeta$ vary as a function of the curvaton mass, with other parameters held fixed. Here the blue curve represents $\eta_B$ variation with $\zeta$ held fixd to its observed value and the red curve represents $\zeta$ variation with mass as $\eta_B$ is fixed to its observed value.


\section{Discussion}
\label{sec:conclusion}

In this {\it Letter}, we have shown that a complex curvaton can also generate the baryon asymmetry
of the universe through the Affleck-Dine mechanism. We have found that this unified construction economically accommodates viable values for both the curvature perturbation and the baryon asymmetry. We emphasize that this is not a new model; the field content and interactions are identical to that of a minimal Affleck-Dine scenario. The key point is that this field serves as a curvaton without requiring any additional modifications. Indeed, if the Affleck-Dine field is light compared to the Hubble rate during inflation, it will necessarily acquire fluctuations and contribute to the total curvature perturbation.

In our treatment, we calculate the baryon asymmetry by solving the classical equations of motion which govern the homogeneous background value of $\chi$. The real and imaginary parts of the curvaton acquire independent fluctuations during inflation and we track their evolution using the $\delta N$ formalism to calculate the total curvature perturbation at the time of $\chi$ decay. Unlike in the minimal curvaton model, the additional degree of freedom allows for enhanced primordial non-Gaussianity.

\section*{Acknowledgments}
We would like to thank Gongjun Choi, Raymond Co, Keisuke Harigaya, and Aaron Pierce for useful discussions. We also thank Gongjun Choi for comments on the manuscript. AI is supported by NSF Grant PHY-2310429, Simons Investigator Award No.~824870, DOE HEP QuantISED award \#100495, the Gordon and Betty Moore Foundation Grant GBMF7946, and the U.S.~Department of Energy (DOE), Office of Science, National Quantum Information Science Research Centers, Superconducting Quantum Materials and Systems Center (SQMS) under contract No.~DEAC02-07CH11359. TO is supported by the University Research Association Visiting Scholars Program and the U.S. Department of Energy under grant No.~DE-SC 0017987. Fermilab is operated by the Fermi Research Alliance, LLC under Contract DE-AC02-07CH11359 with the U.S. Department of Energy. This material is based partly on support from the Kavli Institute for Cosmological Physics at the University of Chicago through an endowment from the Kavli Foundation and its founder Fred Kavli.

\appendix
\section{Second Order Calculation}\label{sec:second}

In this section, we extend the calculation in the text to account for second-order effects required for non-Gaussianities. For each species $i = \chi_R, \chi_I, \gamma$, we take the Ansatz
\be
\zeta_{i} = \zeta_{i}^{(1)}  + \frac{1}{2} \zeta_{i}^{(2)} \,,
\ee
where the $(1)$ and $(2)$ label the first- and second-order perturbations, respectively. Expanding Eq.~(\ref{eq:density contrast relations 1}) to second order yields
\begin{equation}
\begin{split}
    4 & \Omega_\gamma \left[ \frac{1}{2} \left( \zeta_\gamma^{(2)} - \zeta^{(2)} \right) \!+\! 2 \left( \zeta_\gamma^{(1)} - \zeta^{(1)} \right)^2 \right] \\
    & + 3 \Omega_{\chi_R} \left[ \frac{1}{2} \left(  \zeta_{\chi_R}^{(2)} - \zeta^{(2)} \right) + \frac{3}{2} \left( \zeta_{\chi_R}^{(1)} - \zeta^{(1)} \right)^2 \right] \\
    & + 3 \Omega_{\chi_I} \left[ \frac{1}{2} \left(  \zeta_{\chi_I}^{(2)} - \zeta^{(2)} \right) +  \frac{3}{2} \left( \zeta_{\chi_I}^{(1)} - \zeta^{(1)} \right)^2 \right] = 0 \,.
\end{split}
\end{equation}
Solving for the second order perturbation, we find
\begin{equation}\label{eq:zeta2inprogress}
\begin{split}
    \zeta^{(2)} & = f_\gamma \left[ \zeta_\gamma^{(2)} + 4 \left(\zeta_\gamma^{(1)} - \zeta^{(1)}\right)^2 \right] \\
    & + f_{\chi_R} \left[ \zeta_{\chi_R}^{(2)} + 3 \left(\zeta_{\chi_R}^{(1)} - \zeta^{(1)}\right)^2 \right] \\
    & + f_{\chi_I} \left[ \zeta_{\chi_I}^{(2)} + 3 \left(\zeta_{\chi_I}^{(1)} - \zeta^{(1)}\right)^2 \right] \,,
\end{split}
\end{equation}
where we have defined $f_i$ as in Eq.~(\ref{eq:fs}). The expression above can be simplified using the form of the linear order curvature perturbation $\zeta^{(1)}$ in Eq.~(\ref{eq:zeta-final}). Further, we assume that the initial radiation is homogeneous and that the dominant contribution to the curvature perturbation is coming from the curvaton, allowing us to neglect the terms involving $\zeta_\gamma$. This simplifies the result considerably, and we write
\be
\zeta^{(2)} \!=  \! A \zeta_{\chi_R}^{(2)} + B \zeta_{\chi_I}^{(2)} + C \!\left[ \zeta_{\chi_R}^{(1)}  \right]^2\! +\! D \!\left[ \zeta_{\chi_I}^{(1)}  \right]^2\!\! + E \zeta_{\chi_R}^{(1)} \zeta_{\chi_I}^{(1)},~~~
\ee
where 
\be
A &&= f_{\chi_R} \,, \\
    B &&= f_{\chi_I} \,, \\
    C &&= 3 f_{\chi_R} (1 - f_{\chi_R})^2 + 3 f_{\chi_R}^2 f_{\chi_I} \,, \\
    D &&= 3 f_{\chi_I} (1 - f_{\chi_I})^2 + 3 f_{\chi_I}^2 f_{\chi_R} \,, \\
    E &&= -6 f_{\chi_R} f_{\chi_I} ( 2 - f_{\chi_R} - f_{\chi_I}) \,.
\ee
We further assume that each curvaton field evolves linearly between inflationary horizon exit and the beginning of oscillations, which implies $g_a^{\prime\prime} = g_b^{\prime\prime} = 0$ in Eqs. (17) and (18) of Ref. \cite{Assadullahi:2007uw}. This yields the relations
\begin{align}
    \zeta_{\chi_R}^{(2)} = -\frac{3}{2}  \left[ \zeta_{\chi_R}^{(1)}   \right]^2 ~~,~~
    \zeta_{\chi_I}^{(2)} = -\frac{3}{2} \left[ 
    \zeta_{\chi_I}^{(1)}   \right]^2 \,.
\end{align}
Thus, the second order contribution to the curvature perturbation $\zeta = \zeta^{(1)} + \frac{1}{2} \zeta^{(2)}$ is
\begin{equation}
    \zeta^{(2)} = \left(\! C - \frac{3}{2} A \!\right) \left[\zeta_{\chi_R}^{(1)} \right]^2 + \left( \!D - \frac{3}{2} B\! \right) \left[\zeta_{\chi_I}^{(1)} \right]^2 + E \zeta_{\chi_R}^{(1)} \zeta_{\chi_I}^{(1)} \,,
\end{equation}
which we use in our results throughout the paper.

\section{Baryon Isocurvature}\label{app:isocurvature}

Multi-field inflationary models featuring light new degrees of freedom result in isocurvature perturbations on top of the adiabatic spectrum of fluctuations predicted from the inflaton, and so generically have to contend with bounds on the allowed amount of isocurvature coming from CMB observations~\cite{Planck:2018X}. In the minimal (real) curvaton scenario, the perturbations of the curvaton are initially isocurvature in nature, but become adiabatic when the curvaton decays to reheat the universe sometime after inflation. So long as the energy density of the curvaton is sufficiently large at decay, and so long as dark matter, baryons, and radiation are either among these decay products or produced later, the isocurvature bounds are trivially avoided~\cite{Lyth:2001nq,Bartolo:2002vf,Mollerach:1989hu}. 

Because we consider here a complex curvaton, however, we must account for potential isocurvature. Provided the curvaton comes to dominate the energy density prior to decay, the fluctuations in the radial direction $\delta r$ will establish the adiabatic mode after decay. Fluctuations in the angular direction $\delta \theta$ are in general orthogonal, and so constitute isocurvature. These angular fluctuations lead to isocurvature fluctuations in the baryon density, since the baryon number is necessarily produced via dynamics \textit{before} curvaton decay. To avoid this possibility and the stringent constraints on isocurvature, one either needs to tune the initial conditions such that this is vanishing, or compensate the baryon isocurvature with equal and opposite isocurvature in the dark matter. We will describe the first option, which requires less fine tuning overall.

To be more concrete, recall the form of the baryon number density given in Eq.~(\ref{eq:baryon number time evolution}). Since $a^3 r^4 \sim a^{-3} \sim t^{-3/2}$, the integral is saturated near the lower limit $t_i$. One can then approximate production as occurring in a small period of time $\Delta t$ centered about $t_i$~\cite{Rubakov:2017}
\begin{equation}
    n_B \simeq \frac{\lambda' B_\chi}{2 a^3}  a_i^3 r_i^4  \sin(4 \theta_i) \Delta t \,.
\end{equation}
It follows then that at early times,
\begin{equation}
    \frac{\delta n_B}{n_B} \sim 4 \cot(4\theta_i) \delta \theta \,.
\end{equation}
The fluctuations in the angular mode are set during inflation as
\begin{equation}
    \sqrt{\langle \delta \theta^2 \rangle} \simeq \frac{\gamma}{2\pi} \frac{H_i}{r_i} \,,
\end{equation}
where $\gamma$ is an $\mathcal{O}(1)$ factor. Provided that the dark matter and radiation arise from the decays of the curvaton radial mode, $\delta n_B/n_B \simeq \delta \eta/\eta$ sets the entropy perturbation at early times. This is then conserved until horizon re-entry, at which point it contributes to the temperature perturbation as~\cite{Hertzberg:2013mba}
\begin{equation}
    \frac{\delta T}{T} \bigg|_{\rm iso} = - \frac{6}{15} \frac{\Omega_B}{\Omega_m} \frac{\delta \eta}{\eta} \,.
\end{equation}
Substituting in the form of $\delta \eta/\eta$ above, 
\begin{equation}
    \bigg\langle \left( \frac{\delta T}{T} \right)^2 \bigg\rangle_{\rm iso} \simeq \frac{144 \gamma^2}{225 \pi^2} \frac{\Omega_B^2}{\Omega_m^2} \frac{H_i^2}{r_i^2} \cot^2(4 \theta_i) \,.
\end{equation}

Meanwhile, during inflation the radial mode of the curvaton also picks up fluctuations of amplitude $\delta r \sim H_i/2\pi$ such that the dimensionless curvature power spectrum is schematically $\mathcal{P}_\zeta \sim \kappa^2 (H_i/2\pi r_*)^2$, with $r_*$ the value at horizon exit. The adiabatic contribution to the temperature anisotropy is~\cite{Rubakov:2017}
\begin{equation}
    \bigg\langle \left( \frac{\delta T}{T} \right)^2 \bigg\rangle_{\rm adi} \sim \frac{1}{25} \mathcal{P}_\zeta \,,
\end{equation}
with the factor $1/25$ coming from the Sachs-Wolfe effect. We parameterize the fractional contribution of isocurvature to the total CMB temperature anisotropy by 
\begin{equation}
    \alpha_{\mathcal{II}} \equiv \frac{\langle (\delta T/T)^2\rangle_{\rm iso}}{\langle (\delta T/T)^2\rangle_{\rm adi}+\langle (\delta T/T)^2\rangle_{\rm iso}} \,.
\end{equation}
At 95\% CL, \textit{Planck} TT,TE,EE+lowE+lensing data constrains~\cite{Planck:2018X}
\begin{equation}
    \alpha_{\mathcal{II}} < 1.7 \times 10^{-2} \,.
\end{equation}
We have checked that all of our benchmark points of Table~\ref{tab:successful model} satisfy this condition.

\bibliographystyle{utphys3}
\bibliography{biblio}

\providecommand{\href}[2]{#2}\begingroup\raggedright\begin{thebibliography}{10}

\bibitem{Planck:2018vyg}
{\bfseries Planck} Collaboration, N.~Aghanim {\em et~al.}, ``{Planck 2018 results. VI. Cosmological parameters},'' \href{https://dx.doi.org/10.1051/0004-6361/201833910}{{\em Astron. Astrophys.} {\bfseries 641} (2020) A6}, \href{https://arxiv.org/abs/1807.06209}{{\ttfamily arXiv:1807.06209 [astro-ph.CO]}}. [Erratum: Astron.Astrophys. 652, C4 (2021)].

\bibitem{Ellis:2023wic}
J.~Ellis and D.~Wands, ``{Inflation (2023)},'' \href{https://arxiv.org/abs/2312.13238}{{\ttfamily arXiv:2312.13238 [astro-ph.CO]}}.

\bibitem{Rubio:2018ogq}
J.~Rubio, ``{Higgs inflation},'' \href{https://dx.doi.org/10.3389/fspas.2018.00050}{{\em Front. Astron. Space Sci.} {\bfseries 5} (2019) 50}, \href{https://arxiv.org/abs/1807.02376}{{\ttfamily arXiv:1807.02376 [hep-ph]}}.

\bibitem{Lyth:2001nq}
D.~H. Lyth and D.~Wands, ``{Generating the curvature perturbation without an inflaton},'' \href{https://dx.doi.org/10.1016/S0370-2693(01)01366-1}{{\em Phys. Lett. B} {\bfseries 524} (2002) 5--14}, \href{https://arxiv.org/abs/hep-ph/0110002}{{\ttfamily arXiv:hep-ph/0110002}}.

\bibitem{Enqvist:2001zp}
K.~Enqvist and M.~S. Sloth, ``{Adiabatic CMB perturbations in pre - big bang string cosmology},'' \href{https://dx.doi.org/10.1016/S0550-3213(02)00043-3}{{\em Nucl. Phys. B} {\bfseries 626} (2002) 395--409}, \href{https://arxiv.org/abs/hep-ph/0109214}{{\ttfamily arXiv:hep-ph/0109214}}.

\bibitem{Bartolo:2002vf}
N.~Bartolo and A.~R. Liddle, ``{The Simplest curvaton model},'' \href{https://dx.doi.org/10.1103/PhysRevD.65.121301}{{\em Phys. Rev. D} {\bfseries 65} (2002) 121301}, \href{https://arxiv.org/abs/astro-ph/0203076}{{\ttfamily arXiv:astro-ph/0203076}}.

\bibitem{Sakharov:1967dj}
A.~D. Sakharov, ``{Violation of CP Invariance, C asymmetry, and baryon asymmetry of the universe},'' \href{https://dx.doi.org/10.1070/PU1991v034n05ABEH002497}{{\em Pisma Zh. Eksp. Teor. Fiz.} {\bfseries 5} (1967) 32--35}.

\bibitem{Morrissey:2012db}
D.~E. Morrissey and M.~J. Ramsey-Musolf, ``{Electroweak baryogenesis},'' \href{https://dx.doi.org/10.1088/1367-2630/14/12/125003}{{\em New J. Phys.} {\bfseries 14} (2012) 125003}, \href{https://arxiv.org/abs/1206.2942}{{\ttfamily arXiv:1206.2942 [hep-ph]}}.

\bibitem{Cline:2006ts}
J.~M. Cline, ``{Baryogenesis},'' in {\em {Les Houches Summer School - Session 86: Particle Physics and Cosmology: The Fabric of Spacetime}}.
\newblock 9, 2006.
\newblock \href{https://arxiv.org/abs/hep-ph/0609145}{{\ttfamily arXiv:hep-ph/0609145}}.

\bibitem{Affleck:1984fy}
I.~Affleck and M.~Dine, ``{A New Mechanism for Baryogenesis},'' \href{https://dx.doi.org/10.1016/0550-3213(85)90021-5}{{\em Nucl. Phys. B} {\bfseries 249} (1985) 361--380}.

\bibitem{Charng:2008ke}
Y.-Y. Charng, D.-S. Lee, C.~N. Leung, and K.-W. Ng, ``{Affleck-Dine Baryogenesis, Split Supersymmetry, and Inflation},'' \href{https://dx.doi.org/10.1103/PhysRevD.80.063519}{{\em Phys. Rev. D} {\bfseries 80} (2009) 063519}, \href{https://arxiv.org/abs/0802.1328}{{\ttfamily arXiv:0802.1328 [hep-ph]}}.

\bibitem{Hertzberg:2013mba}
M.~P. Hertzberg and J.~Karouby, ``{Generating the Observed Baryon Asymmetry from the Inflaton Field},'' \href{https://dx.doi.org/10.1103/PhysRevD.89.063523}{{\em Phys. Rev. D} {\bfseries 89} no.~6, (2014) 063523}, \href{https://arxiv.org/abs/1309.0010}{{\ttfamily arXiv:1309.0010 [hep-ph]}}.

\bibitem{Cline:2019fxx}
J.~M. Cline, M.~Puel, and T.~Toma, ``{Affleck-Dine inflation},'' \href{https://dx.doi.org/10.1103/PhysRevD.101.043014}{{\em Phys. Rev. D} {\bfseries 101} no.~4, (2020) 043014}, \href{https://arxiv.org/abs/1909.12300}{{\ttfamily arXiv:1909.12300 [hep-ph]}}.

\bibitem{Lin:2020lmr}
C.-M. Lin and K.~Kohri, ``{Inflaton as the Affleck-Dine Baryogenesis Field in Hilltop Supernatural Inflation},'' \href{https://dx.doi.org/10.1103/PhysRevD.102.043511}{{\em Phys. Rev. D} {\bfseries 102} no.~4, (2020) 043511}, \href{https://arxiv.org/abs/2003.13963}{{\ttfamily arXiv:2003.13963 [hep-ph]}}.

\bibitem{Enqvist:2003}
K.~Enqvist and A.~Mazumdar, ``{Cosmological consequences of MSSM flat directions},'' \href{https://dx.doi.org/10.1016/S0370-1573(03)00119-4}{{\em Phys. Rept.} {\bfseries 380} (2003) 99--234}, \href{https://arxiv.org/abs/hep-ph/0209244}{{\ttfamily arXiv:hep-ph/0209244}}.

\bibitem{Enqvist:2009ww}
K.~Enqvist, S.~Nurmi, O.~Taanila, and T.~Takahashi, ``{Non-Gaussian Fingerprints of Self-Interacting Curvaton},'' \href{https://dx.doi.org/10.1088/1475-7516/2010/04/009}{{\em JCAP} {\bfseries 04} (2010) 009}, \href{https://arxiv.org/abs/0912.4657}{{\ttfamily arXiv:0912.4657 [astro-ph.CO]}}.

\bibitem{Enqvist:2009zf}
K.~Enqvist, S.~Nurmi, G.~Rigopoulos, O.~Taanila, and T.~Takahashi, ``{The Subdominant Curvaton},'' \href{https://dx.doi.org/10.1088/1475-7516/2009/11/003}{{\em JCAP} {\bfseries 11} (2009) 003}, \href{https://arxiv.org/abs/0906.3126}{{\ttfamily arXiv:0906.3126 [astro-ph.CO]}}.

\bibitem{Enqvist:2010ky}
K.~Enqvist, A.~Mazumdar, and O.~Taanila, ``{The TeV-mass curvaton},'' \href{https://dx.doi.org/10.1088/1475-7516/2010/09/030}{{\em JCAP} {\bfseries 09} (2010) 030}, \href{https://arxiv.org/abs/1007.0657}{{\ttfamily arXiv:1007.0657 [astro-ph.CO]}}.

\bibitem{Hooper:2023nnl}
D.~Hooper, A.~Ireland, G.~Krnjaic, and A.~Stebbins, ``{Supermassive primordial black holes from inflation},'' \href{https://dx.doi.org/10.1088/1475-7516/2024/04/021}{{\em JCAP} {\bfseries 04} (2024) 021}, \href{https://arxiv.org/abs/2308.00756}{{\ttfamily arXiv:2308.00756 [astro-ph.CO]}}.

\bibitem{Salopek:1990}
D.~S. Salopek and J.~R. Bond, ``Nonlinear evolution of long-wavelength metric fluctuations in inflationary models,'' \href{https://dx.doi.org/10.1103/PhysRevD.42.3936}{{\em Phys. Rev. D} {\bfseries 42} (Dec, 1990) 3936--3962}. \url{https://link.aps.org/doi/10.1103/PhysRevD.42.3936}.

\bibitem{Starobinsky:1982}
A.~A. Starobinsky, ``{Dynamics of Phase Transition in the New Inflationary Universe Scenario and Generation of Perturbations},'' \href{https://dx.doi.org/10.1016/0370-2693(82)90541-X}{{\em Phys. Lett. B} {\bfseries 117} (1982) 175--178}.

\bibitem{Sasaki:1995}
M.~Sasaki and E.~D. Stewart, ``{A General analytic formula for the spectral index of the density perturbations produced during inflation},'' \href{https://dx.doi.org/10.1143/PTP.95.71}{{\em Prog. Theor. Phys.} {\bfseries 95} (1996) 71--78}, \href{https://arxiv.org/abs/astro-ph/9507001}{{\ttfamily arXiv:astro-ph/9507001}}.

\bibitem{Sasaki:1998}
M.~Sasaki and T.~Tanaka, ``{Superhorizon scale dynamics of multiscalar inflation},'' \href{https://dx.doi.org/10.1143/PTP.99.763}{{\em Prog. Theor. Phys.} {\bfseries 99} (1998) 763--782}, \href{https://arxiv.org/abs/gr-qc/9801017}{{\ttfamily arXiv:gr-qc/9801017}}.

\bibitem{Lyth:2004gb}
D.~H. Lyth, K.~A. Malik, and M.~Sasaki, ``{A General proof of the conservation of the curvature perturbation},'' \href{https://dx.doi.org/10.1088/1475-7516/2005/05/004}{{\em JCAP} {\bfseries 05} (2005) 004}, \href{https://arxiv.org/abs/astro-ph/0411220}{{\ttfamily arXiv:astro-ph/0411220}}.

\bibitem{Sasaki:2006kq}
M.~Sasaki, J.~Valiviita, and D.~Wands, ``{Non-Gaussianity of the primordial perturbation in the curvaton model},'' \href{https://dx.doi.org/10.1103/PhysRevD.74.103003}{{\em Phys. Rev. D} {\bfseries 74} (2006) 103003}, \href{https://arxiv.org/abs/astro-ph/0607627}{{\ttfamily arXiv:astro-ph/0607627}}.

\bibitem{Planck:2019kim}
{\bfseries Planck} Collaboration, Y.~Akrami {\em et~al.}, ``{Planck 2018 results. IX. Constraints on primordial non-Gaussianity},'' \href{https://dx.doi.org/10.1051/0004-6361/201935891}{{\em Astron. Astrophys.} {\bfseries 641} (2020) A9}, \href{https://arxiv.org/abs/1905.05697}{{\ttfamily arXiv:1905.05697 [astro-ph.CO]}}.

\bibitem{Assadullahi:2007uw}
H.~Assadullahi, J.~Valiviita, and D.~Wands, ``{Primordial non-Gaussianity from two curvaton decays},'' \href{https://dx.doi.org/10.1103/PhysRevD.76.103003}{{\em Phys. Rev. D} {\bfseries 76} (2007) 103003}, \href{https://arxiv.org/abs/0708.0223}{{\ttfamily arXiv:0708.0223 [hep-ph]}}.

\bibitem{CMB-S4:2016ple}
{\bfseries CMB-S4} Collaboration, K.~N. Abazajian {\em et~al.}, ``{CMB-S4 Science Book, First Edition},'' \href{https://arxiv.org/abs/1610.02743}{{\ttfamily arXiv:1610.02743 [astro-ph.CO]}}.

\bibitem{Planck:2018X}
{\bfseries Planck} Collaboration, Y.~Akrami {\em et~al.}, ``{Planck 2018 results. X. Constraints on inflation},'' \href{https://dx.doi.org/10.1051/0004-6361/201833887}{{\em Astron. Astrophys.} {\bfseries 641} (2020) A10}, \href{https://arxiv.org/abs/1807.06211}{{\ttfamily arXiv:1807.06211 [astro-ph.CO]}}.

\bibitem{Mollerach:1989hu}
S.~Mollerach, ``{Isocurvature Baryon Perturbations and Inflation},'' \href{https://dx.doi.org/10.1103/PhysRevD.42.313}{{\em Phys. Rev. D} {\bfseries 42} (1990) 313--325}.

\bibitem{Rubakov:2017}
V.~A. Rubakov and D.~S. Gorbunov, \href{https://dx.doi.org/10.1142/10447}{{\em {Introduction to the Theory of the Early Universe}: {Hot big bang theory}}}.
\newblock World Scientific, Singapore, 2017.

\end{thebibliography}\endgroup
\end{document}